# Linear dichroism infrared resonance in over-, under-, and optimally-doped cuprate superconductors

A. Mukherjee,[1] J. Seo,[1] M.M. Arik,[1] H. Zhang,[2] C. Zhang,[2] T. Kirzhner,[3] D.K. George,[1] A.G. Markelz,[2] N.P. Armitage,[4] G. Koren,[3] J.Y.T. Wei,[2] and J. Cerne[1]

1 Department of Physics, University at Buffalo, The State University of New York, Buffalo, New York, 14260, USA
2 Department of Physics, University of Toronto, 60 St. George Street, ON M5S1A7, Toronto, Canada
3 Department of Physics, Technion, Haifa 32000, Israel
4 Institute of Quantum Matter, Department of Physics and Astronomy, The Johns Hopkins University, Baltimore, Maryland 21218, USA

By measuring the polarization changes in THz, infrared, and visible radiation over an extended energy range (3-2330 meV), we observe symmetry-breaking in cuprate high temperature superconductors over wide energy, doping, and temperature ranges. We measure the polarization rotation (Re[$\theta_F$]) and ellipticity (Im[$\theta_F$]) of transmitted radiation though thin films as the sample is rotated. We observe a two-fold rotational symmetry in $\theta_F$, which is associated with linear dichroism (LD) and occurs when electromagnetic radiation polarized along one direction is absorbed more strongly than radiation polarized in the perpendicular direction. Such polarization anisotropies can be generally associated with symmetry breakings. We measure the amplitude of the LD signal and study its temperature, energy, and doping dependence. The LD signal shows a resonant behavior with a peak in the few hundred meV range, which is coincident with the mid-infrared optical feature that has been associated with the formation of the pseudogap state. The strongest LD signal is found in under-doped films, although it is also observed in optimally- and over-doped samples. The LD signal is consistent with nematic charge ordering as well as novel magnetoelectric effects.

Even 30 years after its discovery and hundreds of thousands of publications, high-temperature superconductivity (HTS) in the cuprates remains one of the most enigmatic problems in condensed matter physics. The rich behavior of cuprates is evident from their complex phase diagram, where different order parameters, each with their own broken symmetries, compete. A key problem is to understand the pseudogap phase. Whether it is just a precursor to superconductivity, a crossover in behavior, or represents a true long-range ordered phase with its own broken symmetry, is unknown.

Although symmetry-breaking in cuprate HTS has been theoretically predicted long ago [2,3], it is only recently that THz [1] and near infrared (NIR) [4-6] polarization-sensitive spectroscopy measurements have discovered the presence of circular and linear polarization anisotropies in HTS cuprates. Polarization anisotropies can in general be associated with the breaking of discrete symmetries. In NIR (800 meV) measurements [4], a small (10⁻⁶ rad) polarization rotation in reflection (polar Kerr effect) was measured in YBa$_2$Cu$_3$O$_{6+x}$ (YBCO) in the absence of an external applied magnetic field. In YBCO the polarization signal onsets at a temperature slightly below that of the pseudogap transition temperature. As a polar Kerr rotation can only arise in a system that breaks time-reversal symmetry [7,8], the polarization anisotropy should be associated with the onset of a time-reversal symmetry broken phase, possibly one of an unconventional variety [9]. In measurements performed in the THz range (2-6 meV) [1] a large linear dichroism signal, which had its axes of symmetry not aligned along the principal crystal directions was observed in YBCO. The onset temperature of this signal was near the pseudogap temperature $T^*$ and size of the signal itself was larger in under-doped samples. Another study using much higher energy 1.5 eV light [5] demonstrated similar anisotropy, but with little temperature dependence in the linear response and a very prominent order parameter rise of the second harmonic generation (SHG) signal at $T^*$ that can be identified with inversion symmetry breaking. More recently, dc resistance anisotropy with axes of symmetry misaligned from the symmetry axes of the copper oxygen plans was found in $La_{2-x}Sr_xCuO_4$ [10]. A compilation of experiments on symmetry breaking in HTS cuprates can be found in Fig. 3 in Ref. [11]. The origin of these signals is unknown, but the data suggest the presence of a novel broken symmetry phase, possibly nematic stripes [12] or bond-currents [13]. The frequency dependence of these optical anisotropies may be crucial to resolving the microscopic origin of the broken symmetry. For example, since metals typically have a higher conductivity at lower frequencies, one may expect that metallic stripes could produce larger optical anisotropy as the probe frequency decreases.

We have measured polarization anisotropies for a number of HTS cuprate thin films grown on $LaSrAlO_4$ (LSAO) substrates over a broad frequency range. The energy range that we explore (3-2300 meV) spans most of the important energy scales in HTS cuprates, including the pair-breaking energy, the pseudogap energy, the plasma frequency, scattering rates, and the spin gap energy. In our measurements, the real and imaginary components of the Faraday angle ($\theta_F$) correspond to the polarization rotation and ellipticity, respectively, of the transmitted radiation. In the mid-infrared range, linearly polarized, discrete spectral lines produced by $CO_2$ (10.9 – 9.1 μm, 114 – 135 meV) and CO (6 – 5.2 μm, 205-237 meV) gas lasers are transmitted through the sample at near-normal incidence. The change in the transmitted polarization is measured using photoelastic modulation techniques that employ a ZnSe photoelastic modulator, wire grid polarizer, and liquid nitrogen cooled HgCdTe detector. Broadband (100-600 meV) angle measurements are made at room temperature with a Bruker Vertex 70 Fourier transform infrared (FTIR) spectrometer using the same photoelastic modulation techniques. The THz measurements (2-6 meV) were made using THz time domain spectroscopy. The details of experimental techniques are described in Refs. [1,14-16].

The cuprate thin films used in this study were grown epitaxially on LSAO substrates using pulsed laser-ablated deposition (PLD). LSAO substrates are well-suited for these measurements since they are transparent in the THz/infrared/visible and are tetragonal, which eliminates any linear dichroism (LD) or linear birefringence (LB) signals from the substrate. For the films grown at the University of Toronto, a 248 nm KrF excimer laser with a fluence of 2 J/cm$^2$ was used for the deposition. PLD growths of superconducting YBCO films were made using a substrate temperature of 800°C and in 200 mTorr of $O_2$. Superconducting $La_{1.85}Sr_{0.15}CuO_4$ (LSCO) was grown using a substrate temperature of 750°C and in 600 mTorr of $O_2$. As a non-superconducting control, films of the metallic perovskite $LaNiO_3$ (LNO) were also grown on LSAO using a substrate temperature of 800°C and in 200 mTorr of $O_2$. Details of the PLD growth process are given in Refs. [17-19]. The LSCO film was post-annealed in a commercial high-pressure (HP) furnace at 70 atm and 400°C for 12 hours to enhance their superconducting transition temperature ($T_c$) following Ref. [20]. All films have a typical thickness of 100 nm, which was set using material-specific growth rates. The growth rates were determined using atomic force microscopy, by measuring the step heights of unilayer films after they were chemically etched to create a step edge. Films were characterized using x-ray diffraction (XRD) and dc resistance (R) measurements. XRD of YBCO, LSCO, and LNO films were performed using the $\theta - 2\theta$ method and show the expected substrate and film peaks. $R$ vs. $T$ measurements were performed in the van der Pauw configuration. The measurements show superconducting transitions in the YBCO and LSCO films and a metallic $R$ vs. $T$ behavior in the LNO films. Similarly grown YBCO films were also shown to be epitaxially bonded to the substrate using transmission electron microscopy (TEM) in a previous study [17]. Films grown at Technion are described in Ref. [1]. The YBCO films are twinned, with the a and b axes randomly oriented in domains (~1 µm) that are much smaller than the radiation probe beam size (~1 mm).

The complex Faraday angle at zero external magnetic field $B$ was measured for YBCO, LSCO, and LNO thin films grown on LSAO substrates as a function of incoming linear light polarization angle. Figure 1 shows the dependence of Faraday rotation (Re($\theta_F$)) on the YBCO sample orientation measured at 290 K for photon energies of a) 132 meV, b) 228 meV and c) 2330 meV. The panels are arranged with increasing doping towards the bottom of the figure. The samples are named according to whether they are under-doped (UD), optimally-doped (OPD), or over-doped (OVD), followed by the critical temperature and where they were grown, Technion (Tec) or Toronto (Tor). All the HTS samples show clear LD response. The response is sinusoidal as a function of the incident linear polarization with respect to the sample orientation and shows a 180° periodicity, which identifies the rotation as arising from linear polarization anisotropy, not circular polarization anisotropy. The UD70 K Tor shows the largest peak-to-peak amplitude LD signal of 14 mrad at 228 meV. The LD signals in most of the HTS films, especially the UD samples, are maximized near 228 meV. Note that the relative phase of the LD oscillation with respect to the crystal axes is not the same for all the samples, e.g., OPD 90K Tor and UD 50K Tor start near nonzero values unlike the other films. We also observe a frequency-dependent phase shift in some films, e.g., OVD 85K Tec at 132 and 228 meV. There is no observable LD signal from the LSAO substrate nor from a metallic LNO film grown on LSAO as discussed below.

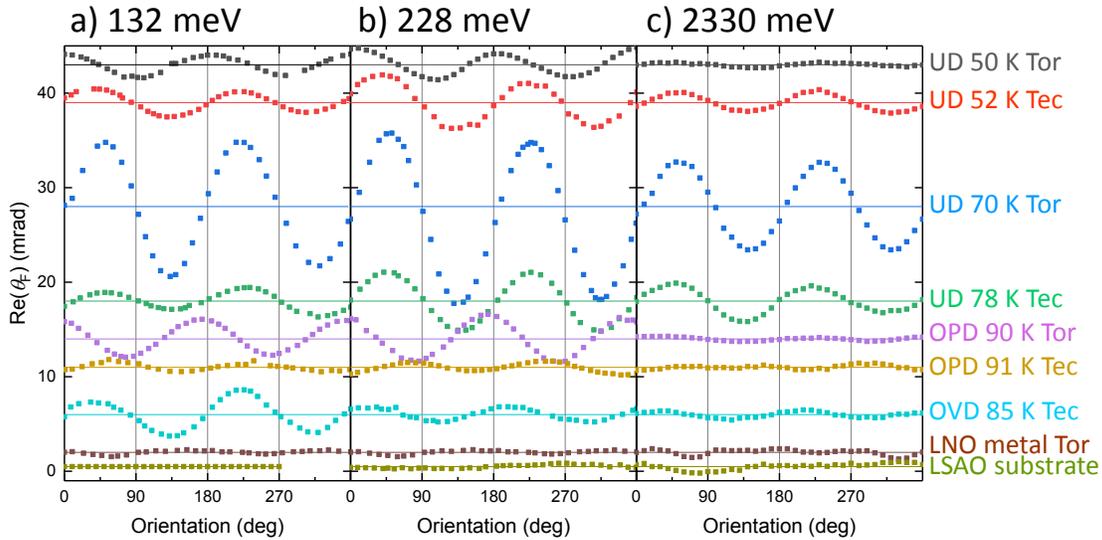

FIG. 1 Sample orientation dependence at 293 K of Faraday angle at three energies a) 132 meV, b) 228 meV, and c) 2330 meV for YBCO films. Samples are listed with increasing doping towards the bottom of the figure. The films grown at Technion and Toronto are labeled Tec and Tor, respectively. The plots are offset vertically, with the zeros for each film given by the colored horizontal lines.

Although prominent in the experiment, the overall scale of the effect can be considered small. If we assign the total polarization changes as due to conductivity anisotropy then the maximum difference in two perpendicular directions is about 1%. In this regard it is important to rule out various artifacts. In related experiments, we have seen the effect of substrate terraces producing infrared optical anisotropy in graphene films on SiC [21]. In those experiments no LD signal was observed in bare insulating SiC. However, when metallic graphene was placed on top of the SiC, the terraces produced anisotropic conductivity due to graphene forming effective "wires" along the terrace edges. This has also been observed in dc transport of graphene on SiC [22]. Effects of this kind need to be excluded and although no LD signal was observed in bare insulating LSAO, we wanted to make sure that unavoidable steps/terraces in LSAO [10] do not produce wire-like conductivity when a metallic film is deposited. To this end, a 100 nm thick film of metallic LNO was deposited on a LSAO substrate. The metal film produced no LD signal, which makes us confident that the LD signal is intrinsic to the cuprate films. We would also point out that a recent dc resistivity study on LSCO thin films has shown a related observation of resistivity anisotropies, the direction of which has no apparent correlation to any steps on the substrate surface [10].

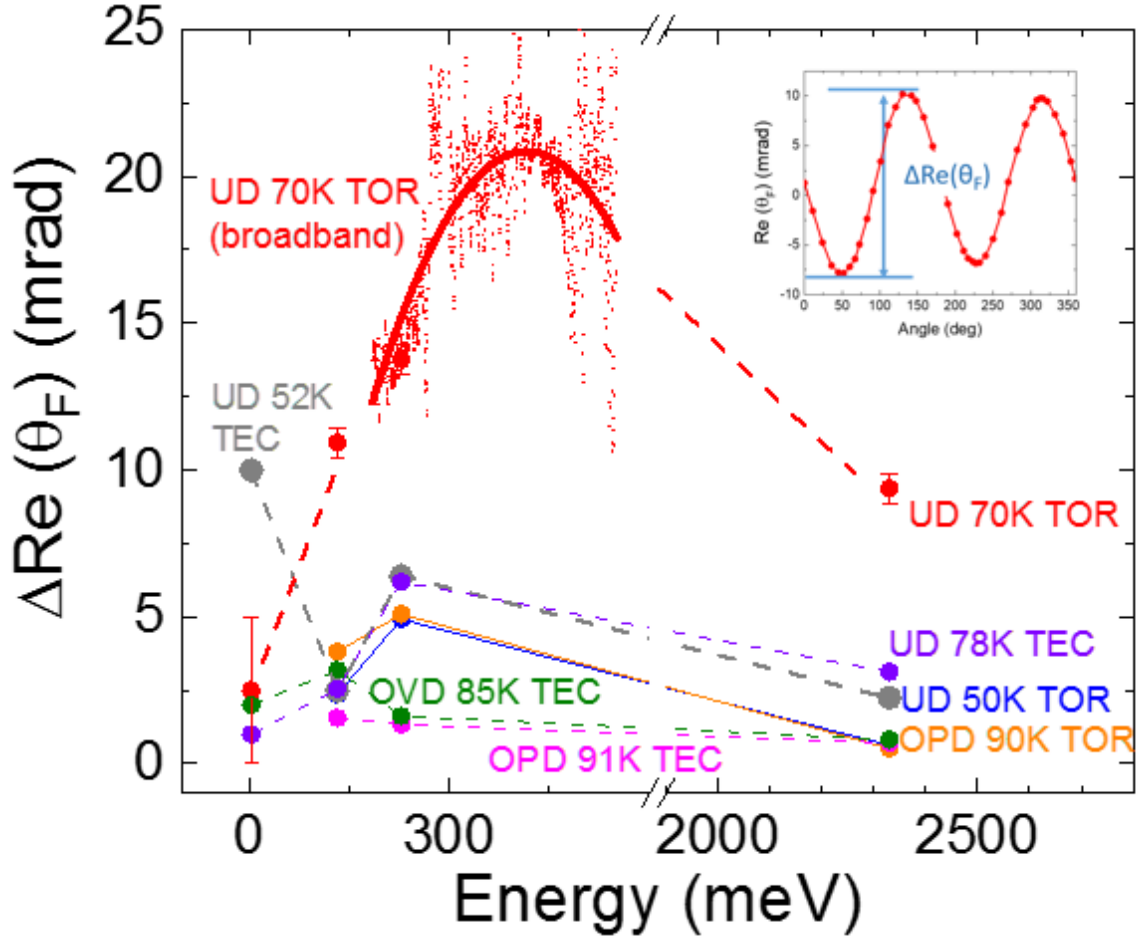

FIG. 2 Energy dependence of Re($\theta_F$) for various YBCO films. The symbols above 50 meV are from measurements using discrete laser emission lines. Symbols near 0 meV are from THz spectroscopy, with UD 78K Tec, OVD 85K Tec, and UD 52K Tec from Ref. [1] and UD 70K Tor measured by the Markelz group. The solid red line for UD 70K Tor from 150-500 meV is a second-order polynomial fit to broadband FTIR measurements. The dashed lines are guides to the eye to connect data from the same film. The inset shows Faraday rotation as a function of sample orientation, from which the amplitude Re($\theta_F$) of the LD signal is determined.

Figure 2 shows the energy dependence from 3-2330 meV of Re($\theta_F$) for YBCO films measured at 290 K and at 0 T. For all the under-doped samples except for UD 52K Tec, the LD signal increases as the probing energy increases from 3 meV towards 300 meV, and then decreases by 2330 meV. The broadband measurements on UD 70K Tor show a peak in the LD signal near 400 meV. For UD 52K Tec, there is also a peak in the signal below 100 meV. The LD signals for optimally-doped (OPD 91 K Tec) and over-doped (OVD 85 K Tec) films were much smaller and decreased with energy above 150 meV. Although OPD 91 K Tec and OVD 85 K Tec show similar frequency dependence, the LD signal for OVD 85 K Tec is over a factor of two larger than that of OPD 91 K Tec at 132 meV. On other hand, the LD signal for the optimally-doped sample OPD 90K Tor is similar in magnitude and energy dependence to the under-doped samples. At the highest energy, 2330 meV, the LD signals for both optimally-doped samples and the over-doped sample are nearly the same and well below the LD signals from the under-doped samples.

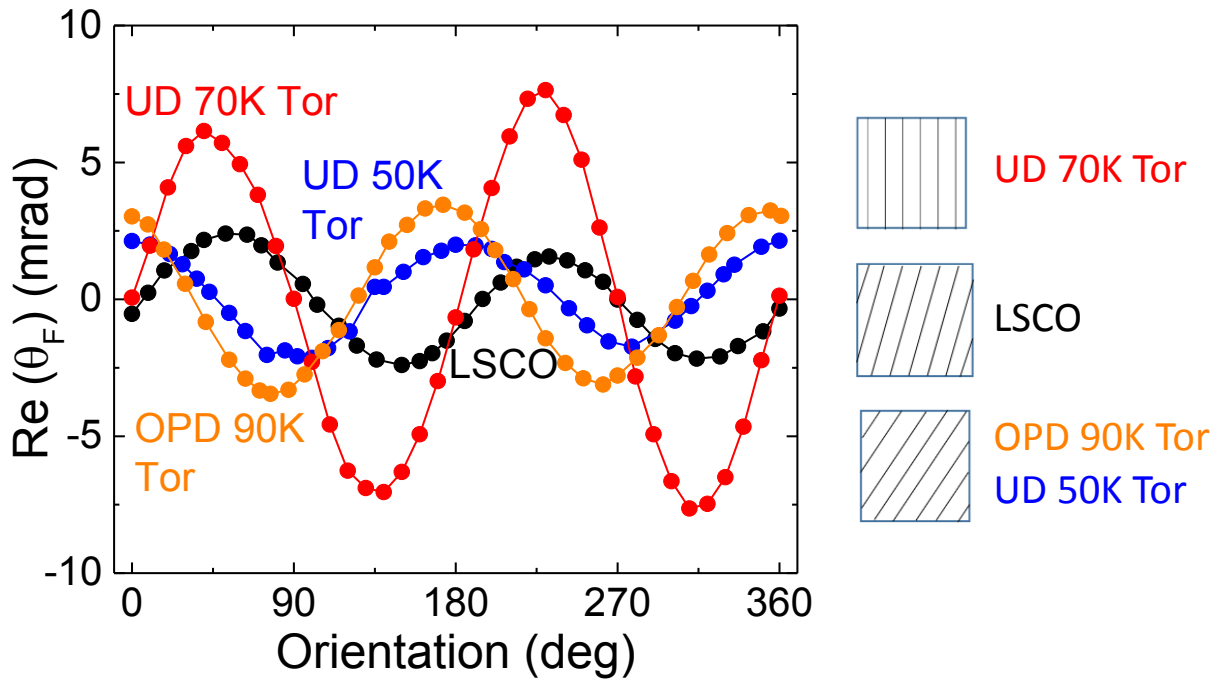

FIG. 3 Phase shifts in the LD signal at 293 K and possible orientations of stripes that are consistent with the LD signals. The UD and OPD films are YBCO with indicated $T_c$s whereas the LSCO film is optimally-doped with a $T_c$ of 34 K.

The LD signal shown in Fig. 1 shows phase shifts and small asymmetry in the lobe amplitudes. If anisotropic conductivity, for example conducting "stripes," is responsible for the LD signal, one may expect that the phase shift depends on the angle that the stripes are oriented with respect to the Cu-O bonds. Figure 3 shows the LD signal measured at 132 meV and the orientation of conducting stripes that would be consistent with the measured signal in this scenario. A lobe asymmetry in the LD oscillations can occur if the angle of incidence of the probing radiation is large (>10°). Therefore in the measurements, great care was taken to minimize the angle of incidence, which we estimate to be below 5°. Since the LSAO substrate is tetragonal, each edge of the substrate is parallel to one of the equivalent crystal axes of the film and we cannot distinguish one edge from the other. However, we can see that in Fig. 3, within such a scenario the stripes could be more parallel to an edge (UD 70 K Tor and LSCO) or closer to 45 degrees from an edge (UD 50K Tor and OPD 90K Tor). Note that even within these two basic groups, small differences in the phase could be observed.

Figure 4(a) shows the temperature dependence of Re($\theta_F$) at 132 meV when a sample is cooled at orientations of ±45° with respect to the polarization of the incident light. The signal changes sign when the sample is rotated by 90°. The difference ΔRe($\theta_F$) between the curves increases as the sample is cooled, which indicates that the LD grows stronger at lower temperature. The LSAO substrate did not show any appreciable temperature dependence. Fig 4(b) tracks ΔRe($\theta_F$) as the samples cool from room temperature to 77 K. The under-doped samples show an increase in ΔRe($\theta_F$) with cooling while the optimally-doped sample shows no temperature dependence and the over-doped sample shows a **decrease** in ΔRe($\theta_F$) with cooling. The over-doped sample behaves in a strikingly different way from the other films, including the optimally-doped sample, with a relatively strong LD signal decreasing with temperature. In THz Faraday measurements [1], this same over-doped film showed a non-monotonic temperature dependence, with two peaks and a dip below 100 K.

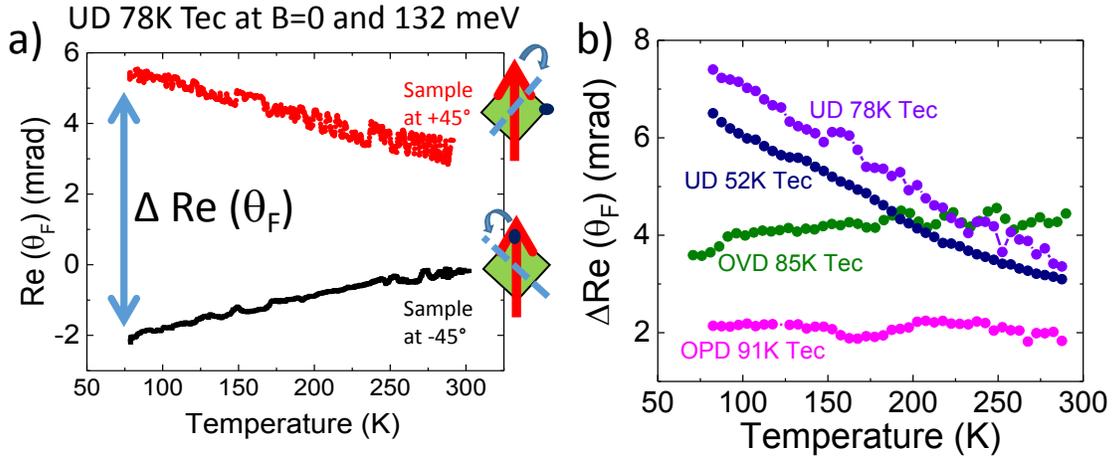

FIG.4 (a) Temperature dependence of Re($\theta_F$) at 132 meV in under-doped YBCO oriented at ±45° with respect to the incident light polarization. (b) Temperature dependence of Δ Re($\theta_F$) for various dopings of YBCO at 132 meV.

Our observations are consistent with the breaking of the naive mirror plane symmetries of the $CuO_2$ planes over a wide range of dopings. In this regard our results are consistent with previous THz spectroscopy measurements, although like the 1.5 eV reflectivity and SHG measurements we find that the symmetry breaking starts at temperatures well above the pseudogap temperature. Clear polarization signals that depend on sample orientation are observed in YBCO and LSCO films from room temperature and down to 77 K without the application of any external magnetic field. We observe strong frequency dependence of the LD signal. The LD signal persists from mid-infrared to visible energies. In most samples, the LD signals is maximized in the few hundred meV energy range. The strongest LD signals tended to be found in under-doped samples, where $T^*$ is highest. Although the samples are multi-domain, one would expect the contributions from crystal structure and the Cu-O chains from different domains to average out, but the strong linear asymmetry signal suggests that the effect is macroscopic and persists across many domains over the large area (~ 0.08 mm$^2$) of the probing beam.

What is the origin of these effects? Both extrinsic and intrinsic possibilities must be considered. For the former it is important to note that the LD signal generally increases with decreased doping and it is important to note that if the signal has its origin in the greater conductivity of Cu-O chains, the trend would be opposite since under-doped samples have less metallic chains. The simplest explanation of the linear dichroism/birefringence structure found here could be in terms of 1D conducting "stripe" or nematic structures [23], where conductivity is high along the wire-like stripes and low in the perpendicular direction. The observation [10] of anisotropy for in-plane dc transport measurements in LSCO is consistent with our observations. The authors in Ref. [10] suggest that the anisotropy is due to electronic nematicity that may not necessarily be aligned with the crystal axes. Among other aspects, these observations appear to rule out more conductive twin boundaries that can exist in YBCO as a source of the anisotropies as they don't exist in LSCO. However, although the previous terahertz polarimetry measurements on hole-doped YBa$_2$Cu$_3$O$_y$ thin films reported the onset of a linear dichroic response near $T^*$ that broke $m_{ac}$ and $m_{bc}$ symmetries [1], it was found later that not only were those symmetries already broken in the crystallographic structure of YBCO crystals above $T^*$, but that SHG experiments showed that the low symmetry of the pseudogap region could not be explained by stripe or nematic type orders alone [5,6].

As pointed out previously [1], another scenario that could account for the interesting tilted optical axes is gyrotropic magnetoelectric birefringence. In such a magnetoelectric effect, magnetic order that breaks both time reversal and inversion symmetries can cause linear birefringence where its zeros deviate from crystalline axes directions. Cr$_2$O$_3$ is the most extensively investigated material with this physics [24]. In principle such an effect can account for a number of unique features of our data including the existence of the tilt. A model based on orbital currents, predicts a magnetoelectric effect that can account for aspects of these data such as the tilt angle and a transition to another loop-current state that breaks mirror symmetry at lower temperature, resulting in a finite Kerr rotation [13]. Recent SHG measurements [6] have found broken symmetries in the cuprates in the pseudogap and regions close to it that are consistent with an odd-parity magnetic order parameter that is broadly consistent with loop current states.

The frequency dependence and maximum we find in the conductivity anisotropy near 300-400 meV may provide an important clue for resolving the microscopic mechanism responsible for the anisotropy. First, a finite frequency maximum appears to most naturally argue against more conductive 1D channels as the source of anisotropy. Metallic response generally peaks at zero frequency, although one can envision (what we consider to be unnatural) scenarios in which there is a larger effective optical mass in one direction, but a larger scattering rate in the other that would give a conductivity anisotropy maximum at finite frequency. In contrast, a finite frequency maximum is a reasonable expectation for a magnetoelectricity as such effects are expected to vanish in the dc limit in a metal. It is important to point out that the frequency range of the maximum is roughly consistent with the frequency range of the mid-infrared absorption [25] that has generally been associated with pseudogap physics. This energy is also close to the two-magnon frequency seen in Raman spectroscopy [26]. On the other hand, the general trend for higher LD as temperature decreases is consistent with higher conductivity, and therefore higher conductivity anisotropy for "wires," as metals are cooled. This temperature dependence is also consistent with bringing the samples into the pseudogap phase as they are cooled. The decrease in the LD signal for the OVD 85K Tec sample as temperature decreases is extremely puzzling and may provide new clues to the symmetry-breaking mechanism(s). To the best of our knowledge, this is the first HTS cuprate that shows a decrease in the symmetry breaking signal as the sample is cooled.

In summary we have observed the temperature and doping dependence of polarization changes to THz, infrared, and visible range (3-2330 meV) light incident on cuprate HTS. We find a two-fold rotational symmetry in the Faraday rotation (Re($\theta_F$)) of the transmitted light and hence this signal is associated with linear dichroism (LD). Such an effect occurs when electromagnetic radiation polarized along one direction is absorbed more strongly than radiation polarized in the perpendicular direction. Such polarization anisotropies can be generally associated with symmetry breakings and in this case the symmetries of the optical anisotropies are generally not directed along crystal axes showing the breaking of crystal mirror symmetries. The LD signal is broadly consistent with nematic charge ordering as well as a novel magnetoelectric effect. However, the signal shows a resonant behavior with a peak in the few hundred meV range, the presence of which may argue against the polarization anisotropy coming from quasi-1D conducting channels and instead may be more consistent with magnetoelectric effects. The peak energy of 300-400 meV is an energy scale generally associated with the pseudogap.

We are indebted to D. Hsieh, S.A. Kivelson, C.M. Varma and L. Zhao for helpful discussions. We gratefully acknowledge support from NSF-DMR 1410599 (JC). AGM and DKG supported by NSF grant MCB 1616529, DOE grant DE-SC0016317 and NIH STTR R41 GM125486. Work in Canada was supported by NSERC, CFI-OIT and the Canadian Institute for Advanced Research.